\documentclass[aps,amsmath,amssymb,twocolumn,prl,superscriptaddress,longbibliography]{revtex4-2}

\usepackage{graphicx}
\usepackage{dcolumn}
\usepackage{bm}
\usepackage{color}
\usepackage{physics}
\usepackage[normalem]{ulem}
\usepackage{amsmath}
\usepackage{comment}
\usepackage{hyperref}
\newcommand{\up}{\uparrow}

\newcommand{\down}{\downarrow}
\renewcommand{\vec}[1]{\textbf{#1}}
\newcommand{\AM}{\mathrm{AM}}

\newcommand{\N}{\mathrm{N}}
\newcommand{\sgn}[1]{\mathrm{sgn}{#1}}

\renewcommand{\S}{\mathrm{S}}

\begin{document}
\title{Anomalous Superfluid Response in Altermagnetic Superconductors}
\author{Christian Wiedemann}
\email{christian.wiedemann@uni-konstanz.de}
\affiliation{Fachbereich Physik, Universität Konstanz, D-78457 Konstanz, Germany}

\author{Danilo Nikoli\'c}
\email{danilo.nikolic@uni-greifswald.de}
\affiliation{Institut f\"ur Physik, Universit\"at Greifswald, Felix-Hausdorff-Straße 6, 17489 Greifswald, Germany}

\author{Matthias Eschrig}
\email{matthias.eschrig@uni-greifswald.de}
\affiliation{Institut f\"ur Physik, Universit\"at Greifswald, Felix-Hausdorff-Straße 6, 17489 Greifswald, Germany}

\author{Wolfgang Belzig} 
\email{wolfgang.belzig@uni-konstanz.de}
\affiliation{Fachbereich Physik, Universität Konstanz, D-78457 Konstanz, Germany}

\date{\today}

\begin{abstract}
 We report on the emergence of the anomalous (paramagnetic) superfluid response in altermagnetic superconductors at arbitrary impurity concentrations. Due to anisotropic gapless superconductivity, altermagnetic superconductors with an out-of-plane Zeeman field display an anisotropic paramagnetic Meißner effect. The effect is strongest for parallel altermagnetic and Zeeman exchange field vectors and in the clean sample. The presence of nonmagnetic impurities leads to isotropisation and, consequently, weakens the effect; however, the paramagnetic response sustains intermediate amounts of impurities in the system. As demonstrated in recent experiments, microwave superfluid stiffness measurements can serve as a sensitive probe of gapless superconductivity. 
 
\end{abstract}

\maketitle
When placed in an external magnetic field, conventional superconductors develop diamagnetic shielding currents that expel the field from the sample interior. This effect is known as the normal (diamagnetic) superfluid response or Meißner effect~\cite{meissnerNeuerEffektBei1933}. However, in recent decades, several superconducting platforms that display the paramagnetic Meißner effect have been proposed~\cite{liParamagneticMeissnerEffect2003,koblischkaParamagneticMeissnerEffect2023}. Those systems include, among others, conventional s-wave superconductors~\cite{geimParamagneticMeissnerEffect1998}, unconventional and multiband superconductors~\cite{braunischParamagneticMeissnerEffect1992,dasilvaGiantParamagneticMeissner2015,parhizgarDiamagneticParamagneticMeissner2021}, topological and multiply connected superconductors~\cite{nielsenParamagneticMeissnerEffect2000,suzukiParamagneticInstabilitySmall2014}, as well as various superconducting hybrid structures~\cite{visaniNovelReentrantEffect1990,fauchereParamagneticInstabilityNormalMetalSuperconductor1999,dibernardoIntrinsicParamagneticMeissner2015,espedalAnisotropicParamagneticMeissner2016,nagyExplanationParamagneticMeissner2016,ouassouPredictionParamagneticMeissner2020a}. 

Recently, the community's attention has been drawn to the discovery of a novel unconventional magnetic ordering termed altermagnetism~\cite{smejkalEmergingResearchLandscape2022,fengAnomalousHallEffect2022}. This unconventional magnetic phase 
follows from the effective decoupling of spin and orbital degrees of freedom, requiring a symmetry description in terms of non-relativistic spin groups and allowing for spin-split but symmetry-compensated bands~\cite{smejkalConventionalFerromagnetismAntiferromagnetism2022,liuSpinGroupSymmetryMagnetic2022}. This enables anomalous Hall effects without net magnetization~\cite{Smejkal2020,fengAnomalousHallEffect2022}, spin-transfer torque~\cite{baiObservationSpinSplitting2022}, and efficient spin-to-charge conversion~\cite{baiEfficientSpintoChargeConversion2023}, while combining antiferromagnetic-like robustness against perturbations with vanishing stray fields and ultrafast dynamics, offering a promising platform for spintronic applications~\cite{baltzAntiferromagneticSpintronics2018}. In connection with mesoscopic superconductivity, various effects have been investigated, including the superconducting proximity effect~\cite{Sun2023,Papaj2023,Sukhachov2024,Chourasia2025,AlamProximitySuperconductivityInAltermagnets2026,delasherasInterplaySuperconductivityAltermagnetism2026,mazanikAbrikosovVorticesAltermagnetic2026}, the symmetry classification of the pairing correlations~\cite{zhangFinitemomentumCooperPairing2024,maedaClassificationPairSymmetries2025,chakrabortyConstraintsSuperconductingPairing2025,fukayaSuperconductingPhenomenaSystems2025b,AlamProximitySuperconductivityInAltermagnets2026,monkmanPersistentSpinCurrentsScAM2026,rasmussenInherentMomentumDependentGapStructureAMS2026}, as well as the normal~\cite{beenakkerPhaseshiftedAndreevLevels2023,Ouassou2023,SunSecondharmonic_2025,Cheng2024,liSpinPolarizedJosephsonSupercurrent2026,darvishiExploringConventionalAnomalous2026} and the anomalous Josephson effect~\cite{Lu2024,alipourzadehAndreevBoundStates2025,fukayaJosephsonEffectOddfrequency2025,sharmaTunableJosephsonDiode2025,boruahFieldFreeJosephson2025,chakrabortyPerfectSuperconductingDiode2024,banerjeeAltermagneticSuperconductingDiode2024,jiangJosephsonDiodeEffect2025a,debnathSpinpolarizationDiodeEffect2025,sahooFieldfreeTransverseJosephson2025,sharmaPwaveMagnetDriven2026, mondalSpinpolarizedAndreevMolecules2026,yangJosephsonDiodeEffect2026,esinJosephsonDiodeSpinValve2026,fuPerfectSpinNonreciprocity2026,niebuhrSpinpolarizedSupercurrentsJosephson2026} in junctions involving unconventional magnets. 

\begin{figure}[t!]
    \centering
    \includegraphics[page=1, width=\linewidth]{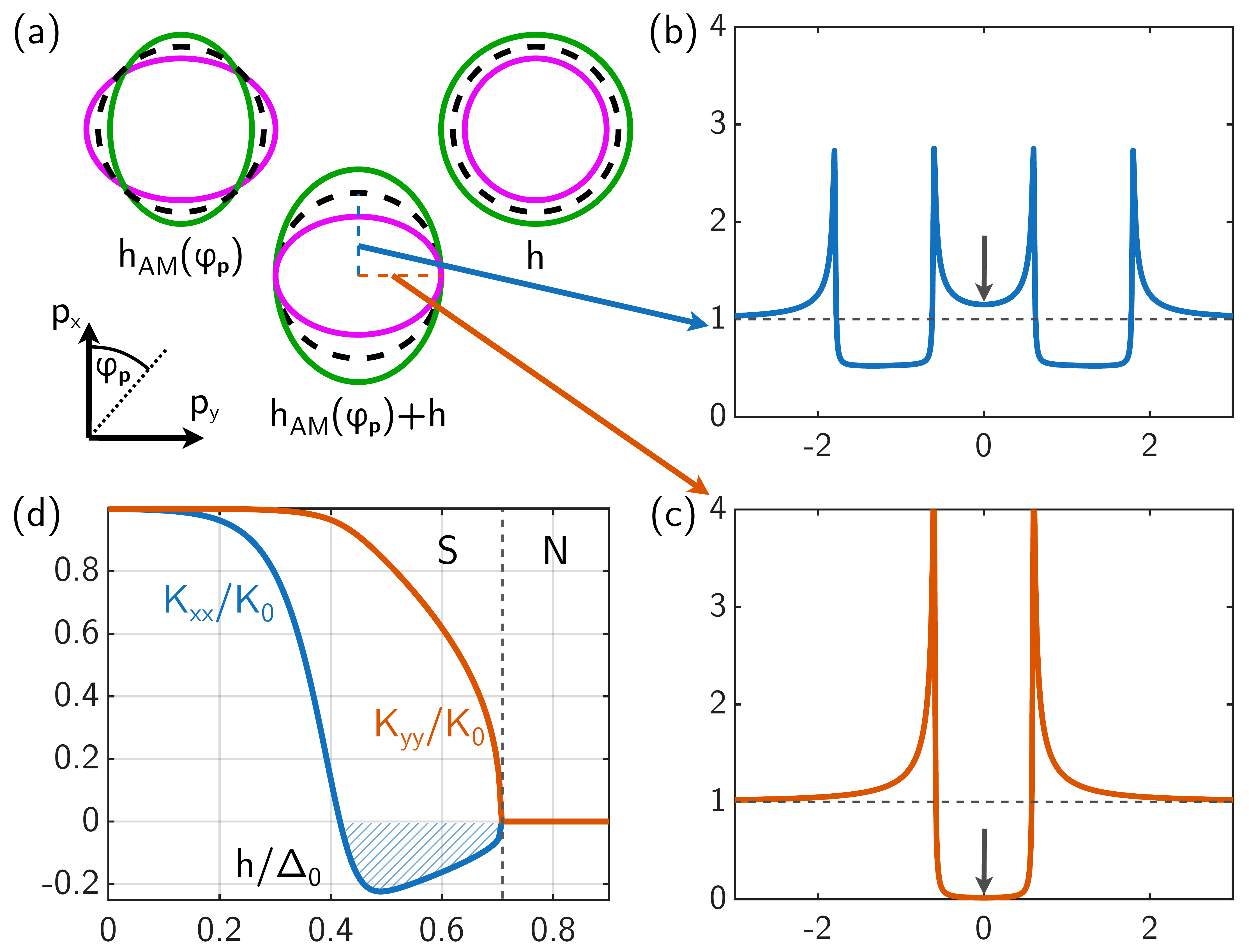}
    \caption{(a) Spin-split Fermi surfaces in a superconductor with altermagnetic exchange field $h_{\AM}$, out-of-plane Zeeman field $\vec{h}=h\vec{e}_z$, and a combination of both (shown for $h_{\AM}=h$). Momentum-resolved DOS, $N/N_0$, along (b) $p_x$ and (c)
    $p_y$, where the former exhibits gapless superconductivity.    
    (d) Meißner kernel as a function of $h$ for $h_\AM=0.6\Delta$ and $T=0.1 T_c$.
    Blue and red lines correspond to $K_{xx}$ and $K_{yy}$, respectively, whilst the off-diagonal components of the $K$ matrix are zero. The blue-shaded area corresponds to the regime in which the system displays the paramagnetic Meißner effect.
    }
    \label{fig:initial}
\end{figure}

In this Letter, we report the emergence of an anomalous (paramagnetic) Meißner effect in altermagnetic superconductors, whose band structure is shown in Fig.~\ref{fig:initial}. Due to the anisotropy of the altermagnetic order, such a system develops gapless superconductivity along certain crystallographic directions. Moreover, the density of states (DOS) can even exceed that of a normal state, as shown in Fig.~\ref{fig:initial}(b), leading to the paramagnetic Meißner effect [see Fig.~\ref{fig:initial}(d)]. To the best of our knowledge, this effect has not been previously discussed. In addition, recent experiments performed on Al/InAs hybrid contacts reported in Ref.~\cite{feyrerEmergenceBogoliubovFermi2026} demonstrated that microwave superfluid stiffness measurements can serve as a sensitive and highly efficient tool for probing gapless superconductivity. 

\textit{Model}--To describe an altermagnetic superconductor, we employ the quasiclassical theory of superconductivity~\cite{larkinQuasiclassicalMethodTheory1969,sereneQuasiclassicalApproachSuperfluid1983,belzigQuasiclassicalGreensFunction1999} described by an Eilenberger equation~\cite{eilenbergerTransformationGorkovsEquation1968}
\begin{equation}\label{eqn:Eilenberger}
 i\hbar\vec{v}_F\cdot\bm{\check\partial}\check{g}_n(\vec{p}_F,\vec{r})+\qty[\check{\mathcal{M}}_n(\vec{p}_F,\vec{r}),\check{g}_n(\vec{p}_F,\vec{r})]=\check{0},
\end{equation}
where~\cite{niebuhrSpinpolarizedSupercurrentsJosephson2026}
\begin{equation}\label{eqn:M}
    \check{\mathcal{M}}_n(\vec{p}_F,\vec{r})=i\omega_n\check{\tau}_z+H(\vec{p}_F)\check{\sigma}_z+\vec{h}\cdot\check{\bm{\sigma}}-\check{\Sigma}_n^\mathrm{imp}-\check{\Delta}(\vec{r}).
\end{equation}
Here, $\bm{\check\partial}(\bullet) = \bm{\nabla}(\check{1}\bullet)-i(e/\hbar)[\vec{A}\check{\tau}_z,\bullet]$ where $\vec{A}$ is the vector potential, $\omega_n=(2n+1)\pi k_BT$ are fermionic Matsubara energies with $n\in\mathbb{Z}$ and temperature $T$, and $H(\vec{p}_F)$ is the altermagnetic exchange field. The latter is anisotropic, depending on the direction of the Fermi momentum $\vec{p}_F$, and we choose the orientation as depicted in Fig.~\ref{fig:initial}(a), i.e., $H(\vec{p}_F)= h_\AM(p_{Fx}^2-p_{Fy}^2)/p_F^2 = h_\AM\cos2\varphi_{\vec{p}}$. The quasiclassical Gor'kov Green's function (GF) has a $4\times 4$ matrix structure in Nambu $\otimes$ spin space~\cite{sereneQuasiclassicalApproachSuperfluid1983,eschrigSpinpolarizedSupercurrentsSpintronics2015}
\begin{equation}\label{eqn:Greens_function}
    \begin{pmatrix}
        g_0\hat{1}+\bm{g}\cdot\bm{\hat{\sigma}} &  (f_0+\bm{f}\cdot\bm{\hat{\sigma}})i\hat{\sigma}_y\\
    - (\tilde{f}_0+\bm{\tilde{f}}\cdot\bm{\hat{\sigma}}^\ast)i\hat{\sigma}_y    & -\tilde{g}_0\hat{1}-{\bm{\tilde{g}}\cdot\bm{\hat{\sigma}^\ast}} 
    \end{pmatrix}{},
\end{equation}
satisfying the normalization condition $\check{g}^2=\check{1}$. The check $(\check\bullet)$ denotes the $4\times 4$ structure in combined space and the tilde ($\tilde{\bullet}$) refers to the particle-hole conjugation operation $\tilde{\mathcal{Q}}_n(\vec{p}_F,\vec{r})=\mathcal{Q}_n^\ast(-\vec{p}_F,\vec{r})$. Since we consider a bulk system, $\bm{\nabla}\check{g}=0$, henceforth, we omit the argument $\vec{r}$ in all quantities. Relevant interactions are included via different self-energies in Eq.~\eqref{eqn:Eilenberger}. Considering the $\vec{e}_z$ direction as a spin quantization axis, the altermagnetic term is given by $\vec{H}\cdot\bm{\check{\sigma}}=H\check{\sigma}_3$, while 
the Zeeman term is $\vec{h}\cdot\check{\bm{\sigma}}$ with $\vec{h}$ pointing in an arbitrary direction, $\vec{h}=h(\sin\theta\cos\phi,\sin\theta\sin\phi,\cos\theta)^T$. Here, $\check{\bm{\sigma}}=\text{diag}(\hat{\bm{\sigma}},\hat{\bm{\sigma}}^\ast)$, where $\hat{\sigma}_i$ are spin Pauli matrices. The electron-impurity scattering is treated within the self-consistent first Born approximation
$\check{\Sigma}_n^\mathrm{imp} = -\frac{i}{2} \hbar\Gamma_\mathrm{imp}\expval{\check{g}_n(\vec{p}_F)}_{\vec{p}_F}$, where $\Gamma_\mathrm{imp}=2\pi N_0 n_\mathrm{imp}|u|^2/\hbar$ is the electron-impurity scattering rate. Superconductivity is included via the BCS self-energy $\check{\Delta}=\mathrm{antidiag}(\hat{\Delta},\tilde{\Delta})$, assuming the singlet pairing, $\hat{\Delta}=\Delta i\hat{\sigma}_y$ and $\tilde{\Delta}=\Delta^\ast i\hat\sigma_y$, and accounting for the self-consistency condition:
\begin{equation}\label{eqn:delta}
    \Delta\ln(\frac{T}{T_c})=-2i\pi k_BT \sum_{n\geq 0} \qty[\expval{f_{0,n}(\vec{p}_F)}_{\vec{p}_F} + \frac{\Delta}{i\omega_n}].
\end{equation}
Calculating observables typically requires averaging over the Fermi surface, denoted by $\expval{\bullet}_{\vec{p}_F}$. Here, we consider a 2D $d$-wave altermagnet with elliptical Fermi surfaces for the two spin bands, whose eccentricities are $\propto\sqrt{h_\AM/E_F}$. Since we examine the interplay between superconductivity and altermagnetism, only weak altermagnetic exchange fields comparable to $\Delta\ll E_F$ are of interest. Consequently, the Fermi surface can be considered as nearly spherical, leading to the simple expression for the Fermi surface averaging: $\expval{\bullet}_{\vec{p}_F} = \int_0^{2\pi}\frac{d\varphi_\vec{p}}{2\pi} (\bullet)$.

\textit{Observables}--Obtaining the quasiclassical GF allows us to express physical observables compactly. For instance, the current density and the DOS read, respectively,
\begingroup
\allowdisplaybreaks
\begin{gather}
    \label{eqn:j_general}
     \bm{j}=\frac{-i\pi e N_0 k_BT}{2}\sum_{n}\Tr\langle\vec{v}_F\check{\tau}_z\check{g}_n(\vec{p}_F)\rangle_{\vec{p}_F} \\
     \label{eqn:dos_general}
    N(\epsilon)=\frac{N_0}{4}\mathrm{Re}\Tr\expval{\check{\tau}_z\check{g}(\epsilon,\vec{p}_F)}_{\vec{p}_F},
\end{gather}
\endgroup
where $N_0=m/(2\pi\hbar^2)$ is the DOS at the Fermi level per spin of a 2D electron gas and $\check{g}(\epsilon,\vec{p}_F)$ is obtained by $i\omega_n\to\epsilon+i\eta$, where $\eta$ is the Dynes parameter~\cite{dynesDirectMeasurementQuasiparticleLifetime1978}. The difference in free energies of the self-consistently treated superconducting state, $\Delta \neq 0$, and the normal state, $\Delta=0$, per unit volume $\mathcal{V}$, is given by~\cite{Burkhardt,burkhardtFFLOstateLayeredSuperconductors1994,eschrigFreeEnergyNonuniform2026}
\begin{align}\label{eqn:free_energy}
    \frac{\Omega_\S - \Omega_\N}{N_0 \mathcal{V}} &= \Delta^2 \log \left( \frac{T}{T_c} \right) +2 \pi k_BT \times\\
    &\times\sum_{n\geq 0}\qty[\int^\infty_{\omega_n} \frac{d\epsilon}{2} \Tr\expval{\check{\tau}_z \qty[\check{g}(\epsilon,\vec{p}_F)-\check{g}_N]}_{\vec{p}_F}+\frac{\Delta^2}{|\omega_n|}],\nonumber
\end{align}
where $\check{g}_N=\sgn{(\omega_n)}\check{\tau}_z$ is the quasiclassical GF for a homogeneous normal metal. Assuming small phase gradients, the superfluid response is calculated as a linear response of the system to the applied field $\vec{A}$~\cite{comment:LL}. Therefore, the solution is sought in the form $\check{g}_n(\vec{p}_F)=\check{g}^{(0)}_{n}(\vec{p}_F) + \check{g}_{n}^{(1)}(\vec{p}_F)$, where $\check{g}^{(0)}_n$ is the solution to Eq.~\eqref{eqn:Eilenberger} for $\vec{A}=0$ and $\check{g}^{(1)}_n$ is a correction linear in $\vec{A}$. Note that due to the altermagnetic field $H(\vec{p}_F)$, even the unperturbed solution $\check{g}^{(0)}_n$ depends on $\vec{p}_F$; however, carrying no current $\langle{\vec{v}_F\check{g}_n^{(0)}(\vec{p}_F)\rangle}_{\vec{p}_F}=\check{0}$. Since $\check{g}_n^{(0)}$ commutes with $\check{\mathcal{M}}_n$ and we only consider linear corrections in $\vec{A}$, Eq.~\eqref{eqn:Eilenberger} simplifies to
\begin{equation}\label{eqn:modified_Eilenberger}
 e \vec{v}_F\cdot\vec{A} \,[\check{\tau}_z,\check{g}^{(0)}_n]+\qty[\check{\mathcal{M}}_n,\check{g}_n^{(1)}]=0.
\end{equation}
This equation we solve using the eigendecomposition, $\check{\mathcal{M}}_n=\check{\mathcal{U}}_n\check{\mathcal{M}}_{\mathrm{diag},n}\check{\mathcal{U}}_n^{-1}$, which leads to
$\left[ \check{\mathcal{M}}_{\mathrm{diag},n} , \check{\mathcal{U}}_n^{-1}\check{g}_{n}^{(1)}\check{\mathcal{U}}_n \right] = -e \vec{v}_F \cdot \vec{A} \check{\mathcal{U}}_n^{-1} \left[ \check{\tau}_z ,\check{g}_n^{(0)} \right] \check{\mathcal{U}}_n$~\cite{SM}.
We find $\check{\mathcal{U}}_n^{-1}\check{g}_{n}^{(1)}\check{\mathcal{U}}_n=e(\vec{v}_F\!\cdot\!\vec{A})\check{\mathcal{X}}_n$, where $\check{\mathcal{X}}_n(\vec{p}_F)$ component-wise reads~\cite{SM}
\begin{equation}\label{eqn:X}
    \check{\mathcal{X}}^{\alpha\beta}_n= 
    \begin{cases}
     \frac{1}{m_\beta - m_\alpha}\big(\check{\mathcal{U}}_n^{-1} \big[ \check{\tau}_z ,\check{g}_n^{(0)}\big] \check{\mathcal{U}}_n\big)_{\alpha\beta},\! & m_\alpha\!\neq\!m_\beta \\
    \qquad\qquad\quad~0\qquad\qquad~~, & m_\alpha\!=\!m_\beta
    \end{cases}.
\end{equation}
Therefore, $\check{g}_n^{(1)}(\vec{p}_F)=e(\vec{v}_F\cdot\vec{A})\check{\mathcal{Y}}_n(\vec{p}_F)$, where $\check{\mathcal{Y}}_n=\check{\mathcal{U}}_n\check{\mathcal{X}}_n\check{\mathcal{U}}_n^{-1}$. Inserting this solution in Eq.~\eqref{eqn:j_general}, we obtain $j_i=-K_{ij}A_j$, where the Meißner kernel reads
\begin{equation}\label{eqn:Meissner_kernel_general}
  K_{ij}\!=\!\frac{ie^2 N_0\pi }{2} k_B T \sum_n\expval {v_{Fi}(\vec{p}_F)v_{Fj}(\vec{p}_F)\Tr\qty[\check{\tau}_z\check{\mathcal{Y}}_n(\vec{p}_F)]}_{{\vec{p}_F}}\!.\!  
\end{equation}
Considering the BCS state in the clean limit yields $\Tr(\check{\tau}_z\check{\mathcal{Y}}_n)=-4i|\Delta|^2/\Omega_n^3$, where $\Omega_n=\sqrt{\omega_n^2+|\Delta|^2}$~\cite{SM}. At zero temperature, this results in the known expression for the Meißner kernel, $K_{0}^{ij}=K_0\delta_{ij}$, where $K_0=e^2 N_0 v_F^2=e^2 \rho_e/m$, with $\rho_e$ being the total electron density (including spin degeneracy)~\cite{abrikosov1963methods,prozorovLondonPenetrationDepth2011}. Henceforth, we express the Meißner kernel in units of $K_0$. For the system at hand, we numerically solve Eq.~\eqref{eqn:Eilenberger} accompanied by the self-consistency condition~\eqref {eqn:delta}, enabling us to compute the observables described above. The results are presented in the following sections. 
\begin{figure}[t!]
    \centering
    \includegraphics[page=1, width=1\linewidth]{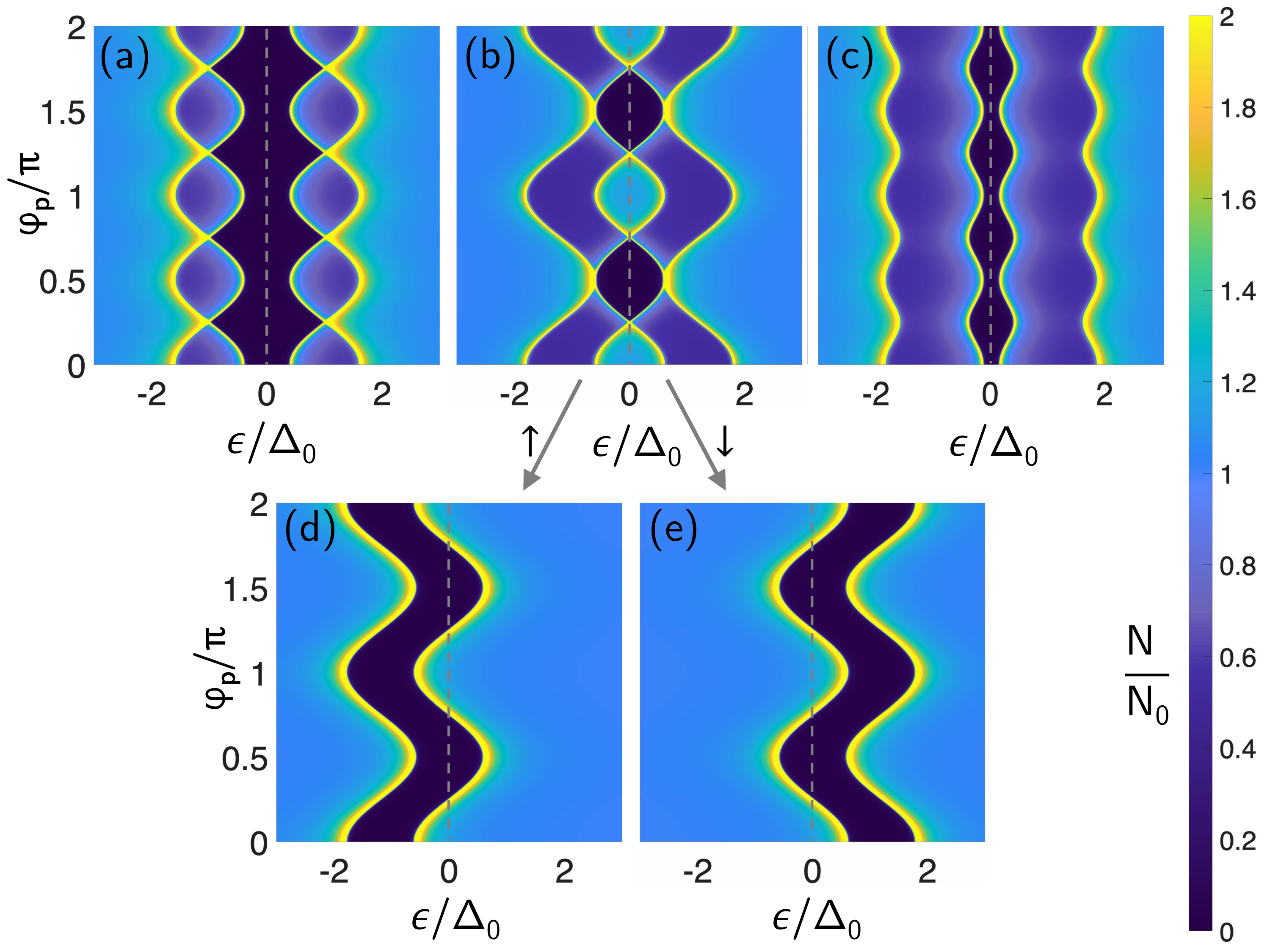}
    \caption{Angle-resolved DOS function for (a) only an altermagnetic exchange field ($h_{\AM}=0.6\Delta_0$, $h=0$), (b) the combination of an altermagnetic field and an out-of-plane Zeeman field, $\vec{h}=h\vec{e}_z,~(h_{\AM}=h=0.6\Delta_0)$, and (c) the combination of an altermagnetic field and an in-plane Zeeman field, $\vec{h}=h\vec{e}_x~(h_{\AM}=h=0.6\Delta_0)$. (c,d) Spin-resolved DOS for the case presented in b). In all panels $T=0.1T_c$.}
    \label{fig:density_of_states}
\end{figure}

\textit{Gapless superconductivity}--As shown in Fig.~\ref{fig:initial}(a), the altermagnetic exchange field causes an anisotropic spin-split Fermi surface compared to the isotropic out-of-plane Zeeman field, $\vec{h}=h\vec{e}_z$. If both are present, they counteract or add up depending on the direction of the Fermi momentum. This leads to a shift in the spin-resolved DOS, $N_{\uparrow/\downarrow}$, which changes along the Fermi surface parametrized by $\varphi_\vec{p}$. For a sufficiently large combined exchange field, $H(\varphi_\vec{p})+h$, this shift exceeds the energy gap, leading to a nonvanishing DOS at the Fermi level that depends on $\varphi_\vec{p}$. Particularly, the gap is maximally filled for $\varphi_\vec{p}=0$ [see Fig.~\ref{fig:initial}(b), while the BCS case is realized for $\varphi_\vec{p}=\pi/2$ [see Fig.~\ref{fig:initial}(c)]. A closer view at the momentum-resolved DOS function, $N(\epsilon,\varphi_\vec{p})$, is shown in Fig.~\ref{fig:density_of_states}.  
\begin{figure*}[t!]
    \centering
    \includegraphics[page=1, width=0.9\textwidth, trim={4.8cm 0 6.24cm 0},clip]{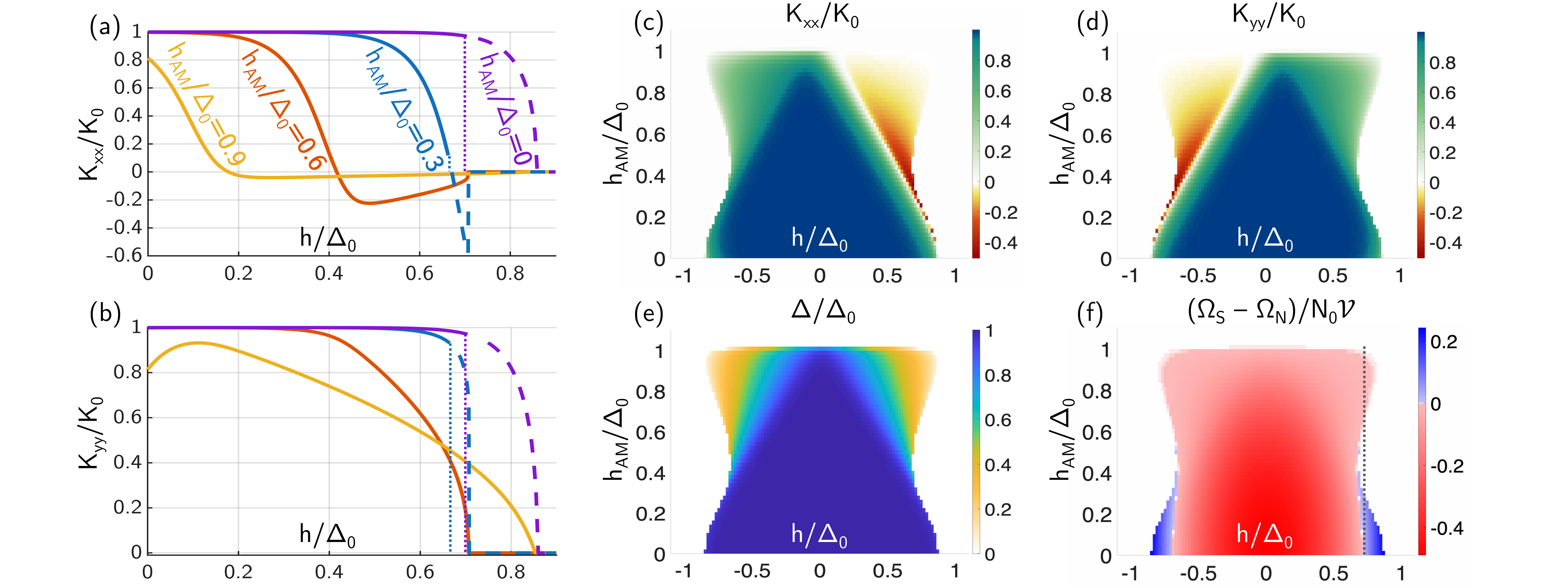}
    \caption{(a) and (b): Meißner kernel's components $K_{xx}$ and $K_{yy}$ as functions of out-of-plane Zeeman $h$ field for various values of $h_\AM$. Dashed lines denote the unstable solution where $\Omega_S > \Omega_N$, while dotted lines indicate the resulting drop to $K_{ij}=0$ for the stable solution $\Delta=0$. (c) and (d): $K_{xx}$ and $K_{yy}$ across the $h_\AM-h$ space. (e) and (f): the corresponding order parameter $\Delta/\Delta_0$ and the free energy difference, $(\Omega_\S - \Omega_\N)/N_0 \mathcal{V}$, where red indicates the region of stable superconductivity. The gray dotted line in (f) indicates where a finite altermagnetic exchange field enables superconductivity beyond the Pauli limit.} In all panels, $T=0.1T_c$.
    \label{fig:kinetic_inductance}
\end{figure*}
Taking into account only the anisotropic altermagnetic field, $\vec{h}=\vec{0}$, causes a $\varphi_\vec{p}$-dependent spin-splitting which shrinks the energy gap [see Fig.~\ref{fig:density_of_states}(a)]; however, no gapless state appears. Adding a Zeeman term parallel to the altermagnetic exchange field shifts the $\up$ and $\down$ spin bands in opposite directions [see the lower panel in Fig.~\ref{fig:density_of_states}\,], so that for $|H(\varphi_\vec{p})+h| >\Delta$ the gap closes in a relatively broad range of values around $\varphi_\vec{p}=n\pi, n\in\mathbb{Z}$, as shown in Fig.~\ref{fig:density_of_states}(b). Furthermore, the averaged DOS does not vanish at the Fermi level, leading to a gapless superconducting state. This scenario is not realized for in-plane fields, $\vec{h}=h\vec{e}_x$, where the energy gap, although shrunken, remains nonzero as long as the system is superconducting [see Fig.~\ref{fig:density_of_states}(c)]~\cite{SM}. 

\textit{Paramagnetic instability and anomalous superfluid response}-- Considering a clean altermagnetic superconductor with Zeeman splitting allows for an analytic solution of Eq.~\eqref{eqn:Eilenberger}. Passing to the frame of reference that diagonalizes the (total) exchange term in spin space~\cite{SM,nikolicSpinresolvedJosephsonDiode2025} leads to $\check{\mathcal{M}}_n=i\omega_n\check{\tau}_z+\mathcal{H}(\varphi_\vec{p})\check{\sigma}_z-\check{\Delta}$, where $\mathcal{H}(\varphi_\vec{p})=\sqrt{h^2\sin^2\theta+(h_\AM\cos2\varphi_\vec{p}+h\cos\theta)^2}$~\cite{chourasiaThermodynamicPropertiesSuperconductor2025}. As a result, the total GF defined in $4 \times 4$ Nambu $\otimes$ spin space factorizes into two $2\times 2$ blocks in Nambu space only, given by $\hat{g}_{n,\sigma}^{(0)}(\vec{p}_F) = \big([\omega_n-i\sigma \mathcal{H}(\varphi_\vec{p})]\hat{\tau}_z-\sigma\Delta\hat{\tau}_y\big)/\Omega_{n,\sigma}(\vec{p}_F)$. Here, $\sigma=\pm$, $\hat{\tau}_i$ are Pauli matrices in Nambu space, $\Delta$ is the magnitude of the superconducting order parameter that is assumed to be real, and $\Omega_{n,\sigma}(\vec{p}_F)=\sqrt{[\omega_n-i\sigma \mathcal{H}(\varphi_\vec{p})]^2+\Delta^2}$. The Meißner kernel $K_{ij}$ is found from the linear response theory, solving Eq.~\eqref{eqn:modified_Eilenberger}, which yields $\hat{g}_{n,\sigma}^{(1)}=(e\vec{v}_F\!\cdot\!\vec{A})(i/2\Omega_{n,\sigma})\hat{g}_{n,\sigma}^{(0)}\big[\hat{\tau}_z,\hat{g}_{n,\sigma}^{(0)}\big]$. Here, we easily identify the $\hat{\mathcal{Y}}_n$ function [see Eq.~\eqref{eqn:X}], which is now defined in Nambu space only as $\hat{\mathcal{Y}}_n=(i/2\Omega_{n,\sigma})\hat{g}_{n,\sigma}^{(0)}\big[\hat{\tau}_z,\hat{g}_{n,\sigma}^{(0)}\big]$. Evaluating the trace $\mathrm{Tr}\big(\hat{\tau}_z\hat{\mathcal{Y}}_n\big)$, we arrive at [see Eq.~\eqref{eqn:Meissner_kernel_general}]:
\begin{equation}
  K_{ij}={e^2 N_0\pi} k_B T\!\sum_{n,\sigma=\pm}\expval {\frac{\Delta^2v_{Fi}(\vec{p}_F)v_{Fj}(\vec{p}_F)}{\Omega_{n,\sigma}(\vec{p}_F)^3}}_{\vec{p}_F}\!.\!
\end{equation}
This expression has a similar form to that of a superconductor with an anisotropic gap~\cite{koganMacroscopicAnisotropySuperconductors2002}.
Expressing explicitly the $K_{xx}$ component, we have $K_{xx}=K_0 \pi k_BT\sum_{n,\sigma}\mathcal{F}_{n,\sigma}$, where
\begin{equation}\label{eqn:F}
    \mathcal{F}_{n,\sigma}=\frac{1}{2}\int\limits_0^{2\pi}\frac{\Delta^2\cos^2\!\varphi\,d\varphi}{{\big([\omega_n-i\sigma \mathcal{H}(\varphi)]^2+\Delta^2\big)^{3/2}}}.
\end{equation}
The component $K_{yy}$ is obtained by replacing $\cos^2\!\varphi\to\sin^2\!\varphi$ in the numerator of Eq.~\eqref{eqn:F}. To obtain a full solution, the above formula should be supplemented by the self-consistency condition~\eqref{eqn:delta}. 

The presence of an out-of-plane Zeeman field, $\vec{h}=h\vec{e}_z$, in an altermagnetic superconductor causes a split between the diagonal components of the Meißner Kernel, $K_{xx}$ and $K_{yy}$, as shown in Fig. \hyperref[fig:initial]{1(d)}, obtained for $h_\AM=0.6\Delta_0$, and $T=0.1 T_c$. Due to the particular orientation of the spatial axes [see Fig.~\ref{fig:initial}(a)], the off-diagonal components, $K_{xy}$ and $K_{yx}$, remain zero. Remarkably, the Meißner response function $K_{xx}$ turns negative for $h\gtrsim0.4\Delta_0$, reaching zero at the transition point to the normal state. In general, a paramagnetic term arises from nonzero DOS at the Fermi level, and in conventional superconductors it is quenched by the energy gap, making them ideal diamagnets~\cite{schrieffer_book,fauchereParamagneticInstabilityNormalMetalSuperconductor1999}. In altermagnetic superconductors, however, the DOS may exceed that of the normal state [see Fig.~\ref{fig:initial}(b)], for which the paramagnetic term exactly cancels the diamagnetic one. Consequently, the paramagnetic term becomes dominant, leading to an anomalous superfluid response. Additionally, the anisotropy of the altermagnetic state leads to the splitting between the $K_{xx}$ [see Fig.~\ref{fig:kinetic_inductance}(a)] and $K_{yy}$ [see Fig.~\ref{fig:kinetic_inductance}(a)] components of the Meißner kernel, where only the former displays a paramagnetic behavior. As shown in Fig.~\ref{fig:kinetic_inductance}(c), the range of $h>0$ for which $K_{xx}$ takes negative values increases with $h_\AM$ due to the enhanced DOS. On the other hand, the $K_{yy}(h)$ function is always positive, referring to a normal (diamagnetic) superfluid response. If the Zeeman field and the altermagnetic field are antiparallel, $h<0$, the effect is mirrored between $K_{xx}$ and $K_{yy}$, as visible in Fig.~\ref{fig:kinetic_inductance}(d). Although the superconducting order parameter is not affected by the direction of the out-of-plane Zeeman exchange field [see Figs.~\ref{fig:kinetic_inductance}(e)], switching its polarization affects the spin polarization. Namely, the two spin bands in our model are defined with respect to the altermagnetic exchange field; therefore, the transformation $h \to-h$ does not simply switch the two bands but shifts them from one another in opposite directions. This can be effectively seen as the swap of $p_x$ and $p_y$, leading to the symmetry $K_{xx}(h)=K_{yy}(-h)$, which is clearly visible in our numerical results presented in Figs.~\ref{fig:kinetic_inductance}(c) and~\ref{fig:kinetic_inductance}(d). 
This paramagnetic Meißner effect does not appear for purely in-plane Zeeman fields since the energy gap is not closed (more details in~\cite{SM}).

\textit{Overall stability of the superconducting state}--Another important aspect of our study is the overall stability of superconductivity in the domain where the anomalous response appears. This is particularly important given that the Pauli paramagnetic limit is substantially different in anisotropic altermagnetic superconductors compared to standard isotropic ferromagnetic superconductors~\cite{chourasiaThermodynamicPropertiesSuperconductor2025}. Figure~\ref{fig:kinetic_inductance}(e) shows the superconducting order parameter $\Delta$ as a function of the altermagnetic field $h_\AM$ and the Zeeman exchange field $h$ (assumed to be along $\vec{e}_z)$. As is known in ferromagnetic superconductors, the solution for $\Delta$ can be multivalued~\cite{makiPauliParamagnetismSuperconducting1964}. In this case, the free energy [see Eq.~\eqref{eqn:free_energy}] has to be minimized, as it determines the realized phase. In Fig.~\ref{fig:kinetic_inductance}(f) we present $\Delta\Omega=(\Omega_\S-\Omega_\N)/(N_0\mathcal{V})$ as a function of $h$ and $h_{\AM}$. We find that the superconducting state is stable, meaning $\Delta\Omega<0$, in most of the $h_{\AM}$-$h$ space (red area), except in the transition regions where $h_{\AM}$ is small compared to $|h|$ (blue area). This region is unstable as $\Delta\Omega>0$, but the order parameter does not vanish. In addition, two solutions for the order parameter are present; however, only one minimizes the free energy corresponding to the first-order phase transition [see the boundary between the red and blue areas in Fig.~\ref{fig:kinetic_inductance}(f)]. This analysis is particularly important as in this region of parameter space, the paramagnetic Meißner effect appears. Note that in our study, we do not consider a possible appearance of the FFLO states~\cite{FFstate,LOstate,chakrabortyConstraintsSuperconductingPairing2025,huUnconventionalSCAltermagneticPolarizedBCSFFLO2025}, as we assume constant $\Delta$. The unstable solution for $\Delta$ affects the superfluid response, and this regime is denoted by the dashed lines in Figs.~\ref{fig:kinetic_inductance}(a) and~\ref{fig:kinetic_inductance}(b). The full free energy consideration gives ground-state solutions denoted by vertical dotted lines.

Since the solutions found in the regions where $\Delta\Omega>0$ are unstable, the critical Zeeman field for $h_{\AM}= 0$ is given by the Pauli limit $|h_{\mathrm{crit}}|=\Delta_0/\sqrt{2}$. Remarkably, a nonzero altermagnetic field can enhance the critical Zeeman field $h_z$ even above the Pauli limit [see the vertical dotted line in Fig.~\ref{fig:kinetic_inductance}(f)]. This effect is due to the counteraction between $h_\AM$ and $h$ that reopens the energy gap. Consequently, the combined critical field is higher than in the cases of the two fields $|h_{\mathrm{crit}}|$ and $h_{\AM,\mathrm{crit}}=\Delta_0$ considered individually. 

\begin{figure}[t!]
    \centering
    \includegraphics[page=1, width=1\linewidth]{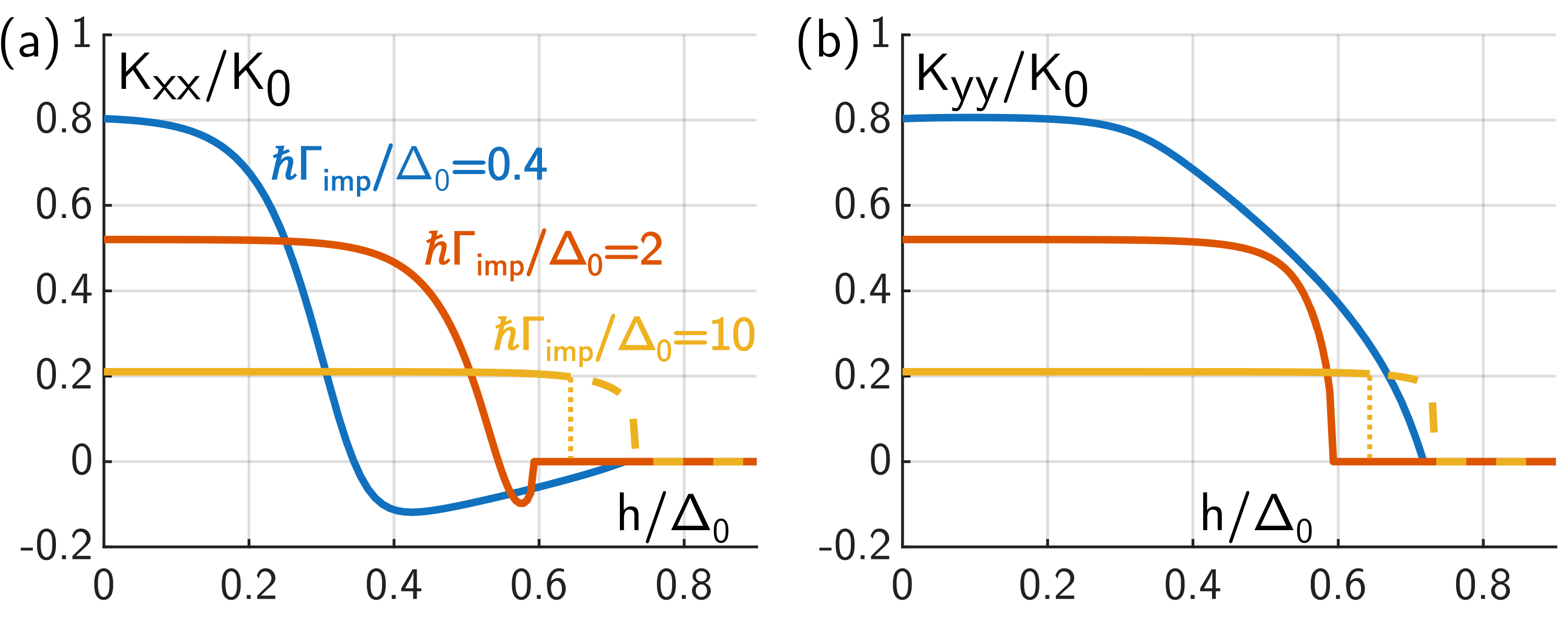}
    \caption{Meißner Kernel components vs. out-of-plane Zeeman field for different impurity scattering rates, $T/T_c=0.1$, and  $h_\AM/\Delta_0=0.8$. The unstable region, $\Delta\Omega>0$, is indicated by dashed lines. The paramagnetic Meißner effect persists at intermediate impurity concentrations, $\hbar\Gamma_\mathrm{imp}\sim\Delta_0$, vanishing in the dirty limit, $\hbar\Gamma_\mathrm{imp}\sim 10\Delta_0.$ }
    \label{fig:impurities}
\end{figure}

\textit{Effect of impurities}-- The results discussed above were obtained for a clean sample, $\Gamma_\mathrm{imp}=0$, as this regime maximizes the paramagnetic Meißner response~\cite{SM}. To include disorder, we account for the impurity self-energy introduced previously. Considering the parallel fields, $\mathcal{H}(\varphi_\vec{p})=H(\varphi_\vec{p})+h$, allows us to express the Meißner kernel in a familiar form $\bar{K}_{xx}=K_0\pi k_B T\sum_{n,\sigma} \bar{\mathcal{F}}_{n,\sigma}$,
where
\begin{equation}\label{eqn:F_imp}
    \bar{\mathcal{F}}_{n,\sigma}=\frac{1}{2}\int\limits_0^{2\pi}\frac{\bar{\Delta}_{n,\sigma}\bar\Delta^\ast_{n,-\sigma}\cos^2\!\varphi\,d\varphi}{{\big([\bar{\omega}_{n,\sigma}-i\sigma \mathcal{H}(\varphi)]^2+\bar{\Delta}_{n,\sigma}\bar\Delta^\ast_{n,-\sigma})^{3/2}}},
\end{equation}
with $\bar{\omega}_{n,\sigma}=\omega_n+(\hbar\Gamma_\mathrm{imp}/2)\expval{g_{n,\sigma}}$ and $\bar{\Delta}_{n,\sigma}=\Delta-i\sigma(\hbar\Gamma_\mathrm{imp}/2)\expval{f_{n,\sigma}}$. Here, $\expval{g_{n,\sigma}}$ and $\expval{f_{n,\sigma}}$ are two additional self-consistent mean fields evaluated from $\expval{g_{n,\sigma}} = \expval{[\bar{\omega}_{n,\sigma}-i\sigma \mathcal{H}(\varphi_\vec{p})]/\bar{\Omega}_{n,\sigma}(\vec{p}_F)}_{\vec{p}_F}$ and $\expval{f_{n,\sigma}} = \expval{-i\sigma\bar{\Delta}_{n,\sigma}/\bar{\Omega}_{n,\sigma}(\vec{p}_F)}_{\vec{p}_F}$ with $\bar{\Omega}_{n,\sigma}(\vec{p}_F)=\sqrt{[\bar{\omega}_{n,\sigma}-i\sigma \mathcal{H}(\varphi_\vec
p)]^2+\bar{\Delta}_{n,\sigma}\bar{\Delta}_{n,-\sigma}^\ast}$~\cite{SM}. The effect of impurities on the Meißner response is presented in Fig.~\ref{fig:impurities}, where the two panels show the $K_{xx}$ and $K_{yy}$ components, as indicated in the figure. As before, only the $K_{xx}$ component displays the anomalous response. As one may expect, the presence of impurities isotropizes the system, weakening the effect. However, the effect can still sustain intermediate impurity concentrations [see the blue and orange lines in Fig.~\ref{fig:impurities}(a)]. As before, special attention should be drawn to the fact that some solutions for the self-consistently calculated gap are unstable. This unstable region is represented by dashed lines in both panels.

\textit{Conclusion}--In summary, we have predicted the emergence of the anomalous anisotropic superfluid response in altermagnetic superconductors with a Zeeman exchange term. This effect is attributed to the gapless superconducting state in such systems. The effect is maximized in a clean sample and for parallel altermagnetic and Zeeman exchange vectors. Due to the nontrivial dependence of the superconducting order parameter on the system parameters, the full solution requires consideration of the free energy, which affects the stability region in the phase diagram. Accounting for impurities, we have shown that the effect sustains intermediate concentrations of $\hbar\Gamma_\mathrm{imp} \gtrsim 2\Delta_0$; however, in the dirty limit, it vanishes due to the isotropization. Our predictions can be tested by measurements of superfluid stiffness that proved to be a sensitive probe of gapless superconductivity, as reported in the recent work of Ref.~\cite{feyrerEmergenceBogoliubovFermi2026}.

\textit{Acknowledgments}––C.W. acknowledges M. Hein for useful discussions. D.N. and M.E. acknowledge funding by the Deutsche Forschungsgemeinschaft (DFG, German Research Foundation) under Project No. 530670387.  W.B. acknowledges support by the Deutsche Forschungsgemeinschaft (DFG; German Research Foundation) via Project No. 465140728 and Project No. 443404566.


\bibliography{references}

\end{document}